\documentclass[]{sig-alternate-10pt}
\usepackage{latexsym}
\usepackage{xspace}
\usepackage{mathptm,graphicx,floatflt,listings}
\usepackage{mathpartir}
\usepackage{todonotes}
\usepackage{listings}
\usepackage{graphics}
\usepackage{algorithm,algorithmic}

\newcommand{\tinysection}[1]{\smallskip {\bf {#1}}.\ }

\usepackage[tight]{subfigure}

\usepackage[small,bf]{caption}

\font\ninett=cmtt8

\font\ninerm=cmr9
\font\ninebf=cmbx9
\font\nineit=cmti9

\font\ninett=cmtt9
\font\ninemi=cmmi9  \skewchar\ninemi='177
\font\ninesy=cmsy9  \skewchar\ninesy='60

\font\mathninerm=cmr9
\font\mathninei=cmmi9
\font\mathninesy=cmsy9

\def\ninepoint{\def\rm{\fam0\ninerm}%
\textfont0=\mathninerm\scriptfont0=\mathsevenrm\scriptscriptfont0=\mathfiverm%
\textfont1=\mathninei\scriptfont1=\mathseveni\scriptscriptfont1=\mathfivei%
\textfont2=\mathninesy\scriptfont2=\mathsevensy\scriptscriptfont2=\mathfivesy%
\textfont3=\tenex
\let\sc=\sevenrm
 \def\it{\fam\itfam\nineit}%
  \textfont\itfam=\nineit
  \def\bf{\fam\bffam\ninebf}%
  \textfont\bffam=\ninebf
  \def\tt{\fam\ttfam\ninett}%
  \textfont\ttfam=\ninett
  \setbox\strutbox=\hbox{\vrule height9pt depth4pt width0pt}%
  \baselineskip=11pt\rm}

\font\matheightrm=cmr8
\font\matheightbf=cmbx8
\font\matheighti=cmmi8
\font\matheightsy=cmsy8
\font\eightit=cmti8
\font\eighttt=cmtt8

\def\eightpoint{%
\textfont0=\matheightrm%
\scriptfont0=\mathsixrm\scriptscriptfont0=\mathfiverm
\textfont1=\matheighti\scriptfont1=\mathsixi\scriptscriptfont1=\mathfivei%
\textfont2=\matheightsy\scriptfont2=\mathsixsy\scriptscriptfont2=\mathfivesy%
\def\rm{\fam0\matheightrm}%
\def\it{\fam\itfam\eightit}%
\textfont\itfam=\eightit%
  \def\bf{\fam\bffam\matheightbf}%
  \textfont\bffam=\matheightbf%
  \def\tt{\fam\ttfam\eighttt}%
  \textfont\ttfam=\eighttt%
  \setbox\strutbox=\hbox{\vrule height7pt depth2pt width0pt}%
 \baselineskip=9pt\rm}

\lstdefinestyle{numbers}
{numbers=left, stepnumber=1, numberstyle=\tiny, numbersep=10pt}

\newfam\smallttfam

\newcommand{\codecf}{\ninett\fam\smallttfam}

\newcount\typett

\newcommand{\cf}[1]{\,{\codecf #1}\,}

\def\startdisplay#1
  {\bgroup\catcode`\`=\active\obeyspaces\obeylines
     \def\i##1{{\eightit ##1}}
     \def\c##1{{\cf ##1}}
     \setlength{\topsep}{0in}
     \setlength{\partopsep}{0in}
     \setlength{\itemsep}{0in}
     \setlength{\parsep}{0in}
     \setlength{\leftmargin}{6.5em}
     \setlength{\rightmargin}{0in}
     \setlength{\itemindent}{0in}
     \baselineskip=9pt
     \codecf\eighttt
     \parskip=0pt plus 1pt\medskip\ifnum\typett>0 \codecf\eighttt\else\it\fi\penalty7000}

\global\def\enddisp{\egroup\catcode`\^^M=5}
 {\obeyspaces\global\let \ }

\newtheorem{theorem}{Theorem}
\newtheorem{lemma}{Lemma}
\newtheorem{definition}{Definition}
\newtheorem{proposition}{Proposition}
\newtheorem{example}{Example}

\newcommand{\barql}{Bar$_{\cal QL}$\@\xspace}

\newcommand{\RULE}[2]{\frac{\begin{array}{c}#1\end{array}}{
                            \begin{array}{c}\\[-2.2mm] #2\end{array}}}

\newcommand{\elide}[1]{}

\DeclareMathAlphabet{\mathpzc}{OT1}{pzc}{m}{it}

\DeclareMathAlphabet{\mathpxbf}{OT1}{pxr}{bx}{n}

\DeclareMathAlphabet{\mathpxit}{OT1}{pxr}{m}{it}

\DeclareMathAlphabet{\mathpxsf}{OT1}{pxss}{m}{n}

\DeclareMathAlphabet{\mathpxtt}{OT1}{pxtt}{m}{n}

\DeclareMathAlphabet{\mathtxbf}{OT1}{txr}{bx}{n}

\DeclareMathAlphabet{\mathtxit}{OT1}{txr}{m}{it}

\DeclareMathAlphabet{\mathtxsf}{OT1}{txss}{m}{n}

\DeclareMathAlphabet{\mathtxtt}{OT1}{txtt}{m}{n}

\newenvironment{nop}{}{}

\def\punto{$\hspace*{\fill}\Box$}

\begin{document}

\title{\barql: Collaborating through Change}
\author{Oliver Kennedy and Lukasz Ziarek\\SUNY Buffalo\\\{okennedy, lziarek\}@buffalo.edu}
\maketitle

\begin{abstract}
Applications such as Google Docs, Office 365, and Dropbox show a growing trend towards incorporating multi-user {\em live collaboration} functionality into web applications.  These collaborative applications share a need to efficiently express shared state, and a common strategy for doing so is a shared log abstraction.
Extensive research efforts on log abstractions by the database, programming languages, and 
distributed systems communities have identified a variety of optimization techniques based on the 
algebraic properties of updates ({\em i.e.}, pairwise commutativity, subsumption, and idempotence).  
Although these techniques have been applied to specific applications and use-cases, to the best of 
our knowledge, no attempt has been made to create a general framework for such optimizations in 
the context of a non-trivial update language.
In this paper, we introduce mutation languages, a low-level framework for reasoning about the 
algebraic properties of state updates, or mutations.
We define \barql, a general purpose 
state-update language, and show how mutation languages allow us to reason about the algebraic 
properties of updates expressed in \barql.
\end{abstract}

\section{Introduction}
\label{sec:introduction}

Over the past several years, many web applications have been released that duplicate and improve on the functionality of desktop applications
({\em e.g.} Google Docs).  A natural consequence of this shift from the desktop to the web is that applications have become more collaborative.  Fully featured word processors, presentation editors, spreadsheets, and drawing programs now exist that allow users to collaboratively edit, view, and annotate documents in ``real-time.''  

Although these {\em collaborative applications} are structured using a client/server model, the core functionality of the application is typically built into the client.  The server's primary role is solely to relay state updates between clients.  In spite of this apparent structural simplicity, collaborative application developers must still expend substantial effort to build scalable and efficient infrastructures for their applications.  

To address this concern, we present the theoretical foundations for a generalized server infrastructure for collaborative applications: Laasie\footnote{Log-As-A-Service InfrastructurE}.  Laasie's primary goal is to encode and replicate application state through a distributed log datastructure.  Clients perform changes to application state by appending them to the log.

The primary motivation for this design is to allow clients to easily recover from link failures ({\em e.g.}, when the host platform changes networks or after it wakes from sleep mode) by maintaining a pointer to the most recent log entry that they have seen.  The server can bring a client up to the most recent state by replaying all log entries that appear after the client's pointer.

Crucially, updates are expressed in the log in terms of {\em intent} rather than {\em effect}.  Below, we introduce and discuss \barql, an update language that can express conditionals and iteration over complex hierarchical datatypes.  Updates expressed in \barql are not evaluated, but rather appended as-is to the log.  This simplifies the semantics of out-of-order appends and makes it easier to express updates as increments 
({\em i.e.}, deltas) rather than fixed write operations ({\em e.g.}, $var := 3$). 

In short, {\em \barql allows the operational semantics of updates to be managed as first class data objects}.

Although an append-only log is a useful high-level abstraction, in practice it becomes necessary to compact the log to bound its size.  For example, a snapshot of the application state can be substituted for all log entries that precede it.  Unfortunately, eliminating all log entries preceding the snapshot also invalidates all clients at states preceding the snapshot as well.  These clients must be (effectively) restarted from scratch, negating the benefits of a log.

In this paper, we present a general framework for reasoning about log updates.  We consider two properties of each rewrite: (1) {\em Correctness}, or whether the rewritten log updates collectively generate a state identical to the original sequence of updates, and (2) {\em Recoverability}, or whether the rewritten log can be used to bring a client at any state up to the most recent state. We then proceed to show how to define {\em incremental} deletion and composition rewrites of the log and provide
realistic ``real-world'' bounds on their behavior. This is accomplished through the definition and use of mutation languages in the following
sections.

The contributions of this paper are as follows:
\begin{enumerate}
\item The design and formalism of {\em mutation languages}, a general framework for reasoning about the correctness and recoverability of log rewrites, and an analysis of the computational complexity of doing so.
\item The construction of mutation languages for composite {\em hierarchical datatypes} derived from mutation languages for simpler primitive types.
\item The formal definition of a log-based update language named {\em \barql}.
\item A reduction from \barql to a composite mutation language, and computational complexity result for computing the correctness and recoverability of log rewrites for \barql
\item An incremental algorithm for identifying candidate log rewrites belonging to two rewrites classes: deletion and composition, with amortized constant time complexity.
\end{enumerate}

\subsection{Roadmap}

Our ultimate goal in this paper is to demonstrate the construction of a practical log rewrite oracle for a non-trivial update language for composite types.  
For any given rewrite of a log, this oracle will determine both the correctness and recoverability of the rewrite. Before defining the oracle we first
define in Section~\ref{sec:semantics-high} a specification of a nontrivial update language (\barql).  
We then use this language to formalize the notion of update logs and log-rewrites, and provide formal definitions of the correctness and recoverability of a log rewrite.

In Section~\ref{sec:semantics-low} we formally define the mutation and mutation language abstractions.  A mutation is simply an expression of
{\em change} and a mutation family is a collection of mutations with properties ({\em e.g.}, commutativity).
 We also identify two binary operations (${\bf merge}$ and ${\bf compose}$) over mutations in a mutation language that 
we will use to simplify the translation of \barql update queries into equivalent mutations.  

Section~\ref{sec:semantics-low:rworacle} outlines the construction of a log rewrite oracle for any mutation language.  
This construction is based on language-specific oracles that evaluate algebraic properties of updates (Commutativity, Subsumption, and Idempotence).

In Section~\ref{sec:reduction} we define a mutation language $\bar{\mathcal L}$ (LBar), and show a reduction from \barql to $\bar{\mathcal L}$.    We provide definitions of ${\bf merge}$, ${\bf compose}$, as well as impractical definitions of the algebraic property oracles.  Using a $\bar{\mathcal L}$, we define a practical set of algebraic property oracles that allow us to construct a log rewrite oracle for \barql.



\section{High Level Semantics}
\label{sec:semantics-high}

\newcommand{\Null}{\mathit{null}}
\newcommand{\dobj}{D_{obj}}
\newcommand{\upd}{{\bf Upd}}
\newcommand{\skel}{\bf Skel}
\newcommand{\tuple}[1]{\left<{#1}\right>}
\newcommand{\condset}[2]{\left\{{#1}\ \left|\ {#2}\right.\right\}}

\newcommand{\merge}{\Leftarrow}
\newcommand{\sng}[2]{\{{#1} := {#2}\}}
\newcommand{\map}[2]{{\bf map}\ {#2}\ {\bf using}\ {#1}}
\newcommand{\agg}[2]{{\bf agg}_{[{#1}]}({#2})}
\newcommand{\mapkey}[2]{{\bf mapkey}\ {#2}\ {\bf using}\ {#1}}
\newcommand{\filter}[2]{{\bf filter}\ {#2}\ {\bf using}\ {#1}}
\newcommand{\ifq}[3]{{\bf if}\ {#1}\ {\bf then}\ {#2}\ {\bf else}\ {#3}}
\newcommand{\id}{{\bf id}}

In this section we introduce \barql, a log-based update language loosely based on the Monad Algebra~\cite{lellahi1997calculus} with unions and aggregates.  Unlike Monad Algebra, which uses sets as the base collection type, \barql uses maps~\footnote{Maps are also popularly referred to as hashes, dictionaries, or lookup tables.} and has weaker type semantics along the lines of \cite{buneman1997semistructured}.  Furthermore, \barql is intentionally limited to operations with linear computational complexity in the size of the input data; neither the pairwith nor cross-product operations of Monad Algebra are included.  In our domain, this is not a limitation, as the server is acting primarily as a relay for state.  Full cross-products can be transmitted to clients more efficiently in their factorized form, and each client is expected to be capable of computing cross products locally\footnote{Joins are an area of concern however, and future work will consider extensions to \barql for this purpose.}.

The domains and grammar 
for \barql are given in Fig.~\ref{fig:domains}. We use $C$ to range over constants, $p$
over primitives (strings, integers, floats, and booleans), $k$ over keys, $Q$ over queries, $\tau$ over types, $v$ over values of type $\tau$, and $\theta$ over binary operations over primitive types.  The type $\tau$ operated over by \barql queries is identical to the labeled trees of \cite{buneman1997semistructured}, and is equivalent to  unstructured XML or JSON. 
Values are either of primitive type, null, or collections (mappings from $k$ to $\tau$).
Note that collections are total mappings; for instances, a singleton can be defined as the collection where all keys except one map to the $\Null$ value.
By convention, when referring to collections we will implicitly assume the presence of this mapping for all keys that are not explicitly specified in
the rules themselves.

We formalize
\barql in Fig.~\ref{fig:barql} in terms of a big-step operational semantics. Order of evaluation
is defined by the structure of the rules.

\begin{figure*}[t]
\[
\begin{array}{lcllcl}
c    & \in & \mathit{Constant} : \rightarrow p & k    & \in & \mathit{Key}\\
p    & \in & \mathit{Primitive}                & Q    & \in & \mathit{Query} : \tau \rightarrow \tau\\
v    & \in & \mathit{Value}                    & \tau & \in & \mathit{Type} : p \; |\; \{k_i \rightarrow \tau_i \} \; |\; \Null\\
\theta & \in & \mathit{Binary Op} 
\end{array}
\begin{array}{lcl}
Q & := & Q.k \;|\; Q \merge Q \;|\; \map{Q}{Q} \;|\; Q\ {\bf op}_{[\theta]}\ Q\\ 
  & |  &  \agg{\theta}{Q} \;|\; \agg{\merge}{Q} \;|\; \filter{Q}{Q}\\
  & |  & \ifq{Q}{Q}{Q} \;|\; Q \circ Q \;|\; {\bf c} \;|\; {\bf null} \;|\; {\bf \emptyset}
\end{array}
\]
\caption[]{Domains and grammar for \barql.}
\label{fig:domains}.
\end{figure*}

In \barql queries are monads, structures that represent computaiton. Reducing the query corresponds to evaluting the computation
expressed by that query.  
The rules for {\em PrimitiveConstant}, {\em Null} and {\em EmptySet} all defined operations take an input value 
and produce a constant value reguardless of input.  The rule {\em PrimitiveConstant} produces a primitive
constant $c$, the rule {\em Null} produces the $\Null$ value, and the rule {\em EmptySet} produces
a empty set. We define an empty set a collection that is a total mapping where
all keys map to the $\Null$ value, represented as: $\{* \rightarrow \Null\}$. 
 The {\em Identity} operation passes through the input value unchanged.
{\em Subscripting} and {\em Singleton} are standard operations.  
In comparison to Monad Algebra, these operations correspond to not only the singleton operation over sets, 
but also the tuple constructor and projection operations.  Because collection elements are identified by keys, we can 
reference specific elements of the collection in much the same way as selection from a tuple.  


The most significant way in which \barql differs from Monad Algebra is its use of the {\em Merge} operation ($\merge$) instead of set union ($\cup$).  $\merge$ combines two sets, overwriting undefined entries (keys for which the collection maps to $\Null$) with their values from the other collection.    
$$(\sng{A}{1} \merge \sng{B}{2})(\Null) = \{A \rightarrow 1, B \rightarrow 2\}$$
If a key is defined in both collections, the right collection takes precedence.
$$(\sng{A}{1} \merge \sng{A}{2})(\Null) = \{A \rightarrow 2\}$$

The merge operator can be combined with singleton and identity to define updates to collections:
$$(\id \merge \sng{A}{3})(\{A \rightarrow 1, B \rightarrow 2\}) = \{A \rightarrow 3, B \rightarrow 2\}$$

Subscripting can be combined with merge, singleton, and identity to define point modifications to collections.
\begin{multline*}
(\id \merge \sng{A}{(\id.A \merge \sng{B}{2})})(\{A \rightarrow \{C \rightarrow 1\}\}) \\
= \{A \rightarrow \{B \rightarrow 2, C \rightarrow 1\}\}
\end{multline*}

Primitive binary operators are defined monadically with operation {\em PrimBinOp}, and include basic arithmetic, comparisons, and boolean operations.  These operations can be combined with identity, singleton, and merge to define updates.  For example, to increment $A$ by 1, we write
$$\{\id \merge \{A := \id.A + 1\}\}(\{A \rightarrow 2\}) = \{A \rightarrow 3\}$$

\barql provides constructs for mapping, flattening and aggregation.  The {\em Map} operation is analogous to its definition in Monad Algebra, save that key names are preserved.  The {\em Flatten} operation is also similar, except that it uses $\merge$, instead of $\cup$ as in Monad Algebra.  The {\em PrimitiveAggregation} class of operators defines aggregation using any closed binary operator $\theta$ operating over over primitive type.  

To increment all children of the root by 1 we write:
$$(\map{(\id + 1)}{\id})(\{A \rightarrow 1, B \rightarrow 2\}) = \{A \rightarrow 2, B \rightarrow 3\}$$

To increment the child $C$ of each child of the root by 1, we write
\begin{multline*}
(\map{(\id \leftarrow \sng{C}{\id.C + 1})}{\id})(\\
\{A \rightarrow \{C \rightarrow 1\}, B \rightarrow \{C \rightarrow 2, D \rightarrow 1\}\} \\
) = \{A \rightarrow \{C \rightarrow 2\}, B \rightarrow \{C \rightarrow 3, D \rightarrow 1\}\}
\end{multline*}

Finally, \barql supports {\em Conditionals} and {\em Filtering}, as well as {\em Composition} of queries.

\begin{figure*}[t]

\[
\begin{array}{lclc}
\mathit{Primitive Constant} &
\RULE{}
     {\widehat{c}(v) \mapsto c} &
\mathit{Null} &
\RULE{}
     {\widehat{\Null}(v) \mapsto \Null}\\
\end{array}
\begin{array}{lclc}
\mathit{Empty Set} &
\RULE{}
     {\widehat{\emptyset}(v) \mapsto \{* \rightarrow \Null\}} &
\mathit{Identity} &
\RULE{}
     {\id(v) \mapsto v}\\
\end{array}
\]
\[
\begin{array}{lclc}
\mathit{Subscripting} &
\RULE{Q(v) \mapsto \{..., k \rightarrow r, ...\}}
     {(Q.k)(v) \mapsto r}&
\mathit{Singleton} &
\RULE{Q(v) \mapsto r}
     {\sng{key}{Q}(v) \mapsto \{k \rightarrow r, * \rightarrow null\}}
\end{array}\\
\]
\[
\begin{array}{lc}
\mathit{Merge} &
\RULE{Q_1(v) \mapsto \{k_i \rightarrow r_i\} \;\;\;\;
      Q_2(v) \mapsto \{k_j \rightarrow r_j\}}
     {(Q_1 \widehat{\merge} Q_2)(v) \mapsto
         \{k \rightarrow r\ |\ (k = k_i = k_j) \wedge (((r = r_i) \wedge (r_j = \Null)) \vee ((r = r_j) \wedge (r_j \neq \Null)))\}}
\end{array}\\
\]
\[
\begin{array}{lclc}
\mathit{Map} &
\RULE{Q_{coll}(v) \mapsto \{k_i \rightarrow v_i\} \\
      Q_{map}(v_i) \mapsto r_i}
     {(\map{Q_{map}}{Q_{coll}})(v) \mapsto \{k_i \rightarrow r_i\ |\ v_i \neq \Null\}}&
\mathit{PrimBinOp} &
\RULE{Q_1(v) \mapsto r_1:p \;\;\;\;
      Q_2(v) \mapsto r_2:p \\
      \theta \in \{+,*,-,/,=,{\bf AND},{\bf OR},\neq,<,\leq,>,\geq\}}
     {(Q_1 \widehat{\theta} Q_2)(v) \mapsto r_1 \theta r_2 }
\end{array}
\]
\[
\begin{array}{lclc}
\mathit{Flatten} &
\RULE{Q_{coll}(v) \mapsto \{k_i \rightarrow v_i\}}
     {(\agg{\merge}{Q_{coll}})(v) \mapsto (v_0 \merge v_1 \merge \ldots)} &
\mathit{Primitive Aggregate} &
\RULE{Q_{coll}(v) \mapsto \{k_i \rightarrow v_i\}}
     {(\agg{\theta}{Q_{coll}})(v) \mapsto (((v_0 \theta v_1) \theta v_2) \theta\ldots)}
\end{array}\\
\]
\[
\begin{array}{lc}
\mathit{IfThenElse} &
\RULE{Q_{cond}(v) \mapsto true \;\;\;\;
      Q_{then}(v) \mapsto r_{then}}
     {(\ifq{Q_{cond}}{Q_{then}}{Q_{else}})(v) \mapsto
         r_{then}}
\RULE{Q_{cond}(v) \mapsto false \;\;\;\;
      Q_{else}(v) \mapsto r_{else}}
     {(\ifq{Q_{cond}}{Q_{then}}{Q_{else}})(v) \mapsto
         r_{else}}
\end{array}
\]
\[
\begin{array}{lclc}
\mathit{Filter} &
\RULE{Q_{coll}(v) \mapsto \{k_i \rightarrow v_i\}\;\;\;\;
      Q_{cond}(v_i) \mapsto t_i }
     {(\filter{Q_{cond}}{Q_{coll}})(v) \mapsto \{ k_i \rightarrow v_i\ |\ t_i\wedge v_i \neq \Null\}} &
\mathit{Composition} &
\RULE{Q_1(v) \mapsto r_1 \;\;\;\;
      Q_2(r_1) \mapsto r_2}
     {(Q_1 \circ Q_2)(v) \mapsto r_2}\\
\end{array}
\]

\caption[]{A formal operational semantics for \barql.}
\label{fig:barql}
\end{figure*}

\newcommand{\readset}{\mathcal R}
\newcommand{\writeset}{\mathcal W}

\newcommand{\mut}{\mathcal M}
\newcommand{\sketch}{\mathbb M}
\newcommand{\nodes}[1]{\mathcal N({#1})}

\newcommand{\mutlang}{\mathcal L}
\newcommand{\coll}[1]{{#1}_{{\bf c}}}
\newcommand{\colltype}{\coll\tau}
\newcommand{\compmut}{\coll\mut}
\newcommand{\mutsketch}{\mathbb L}
\newcommand{\mutmerge}{{\bf merge}}
\newcommand{\mutcmp}{{\bf compose}}
\newcommand{\oraclesub}{\mathcal S}
\newcommand{\oraclecomm}{\mathcal C}

\section{Mutation Languages}
\label{sec:semantics-low}
We will now temporarilly step back from \barql in order to refine our understanding of {\em update logs}.  At its simplest, an update log encodes a state value as a sequence of state mutations applied iteratively, first to a default ``empty" state, and then to the output of the prior transformation.  

If the fundamental primitive of an update log is the state transformation, then the fundamental operation is composition of state mutations.  As a basis for reasoning about the safety properties of changes to this log, we begin with an outline for simple algebras over the composition of state mutations.  

\begin{definition}
A mutation is an arbitrary transformation $M : \tau \mapsto \tau$ mapping values of some state type $\tau$ to new values of the same type.  
A mutation may be parameterized by an set of additional values $R$.  
We write such a mutation as $M_R(v)$.  We say the mutation $M_R(v)$ is:
\begin{itemize}
\item \ldots {\bf destructive} if $M_R$ is independent of $v$.
\item \ldots {\bf idempotent} if $\forall v, R : M_R(v) \equiv M_R(M_R(v))$
\end{itemize}
\end{definition}

\begin{example}
\label{ex:toymutations}
Consider an application that encodes its state as a single integer ({\it i.e.}, $\tau = \mathbb Z$).  Such an application might employ the two mutations ``replace by 0'', and ``increment by 1'':
\begin{center}
\begin{tabular}{lr}
$M_{\tt := 0}(x) \mapsto 0$ &
$M_{\tt ++}(x) \mapsto x + 1$
\end{tabular}
\end{center}
The replace operation is both destructive and idempotent.  The increment operation is neither.  

We can use parameters to create families of mutations.  For example, we can use a single parameter $Y$ to define a family of mutations ``replace by Y'' ($M_{\tt:= Y}$), or ``increment by Y'' ($M_{\tt +=Y}$).  
\end{example}

Having defined mutations in the abstract as functions, we can now formally define the abstract composition of mutations as simple left-first function composition.
$$(M \circ M')(x) \equiv M'(M(x))$$

\begin{proposition}
\label{compisassoc}
Composition is associative.
\end{proposition}

\proof{By Equivalence
$$((M \circ M') \circ M'')(x) \equiv M''(M'(M(x))) \equiv (M \circ (M' \circ M''))(x)$$
\punto}

We can define a composition algebra for any set of mutations $\vec M$ with identical kinds.  We consider two properties in this algebra: (1) pairwise commutativity and (2) subsumption.  Unlike the traditional algebraic notion of commutativity, we consider only the pairwise commutativity of individual mutations.  That is, instead of saying that $\circ$ is commutative, we say that $M$ and $M'$ {\em commute} iff $(M \circ M') \equiv (M' \circ M)$.  Subsumption is also defined pairwise; we say that $M'$ {\em subsumes} $M$ iff $M \circ M' \equiv M'$.

\begin{definition}
\label{def:mutlang}
A mutation language $\mutlang$ is the 4-tuple: \\
$\tuple{\tau, \vec M, \oraclesub, \oraclecomm}$
consisting of:
\begin{enumerate}
\item A state type $\tau$
\item A set of mutations $\vec M$ of kind $\tau \mapsto \tau$.  This set must include the identity mutation $\id(x) \mapsto x$.
\item A binary relation $\oraclesub(M, M')$ that holds if $M$ is subsumed by $M'$.
\item A symmetric binary relation $\oraclecomm(M, M')$ that holds if $M$ commutes with $M'$.
\end{enumerate}
We will use the shorthand $\oraclesub(M) \equiv \oraclesub(M, M)$ to denote the unary idempotence relation.
\end{definition}

A mutation language encapsulates the composition algebra for a specific set of mutations, together with a set of rules for determining idempotence, pairwise commutativity, and subsumption on mutations in the language.  

\begin{example}
\label{ex:toylanguage}
On simple mutation languages, these properties can be determined quite efficiently.  For the mutation language defined from the mutation language families in Example \ref{ex:toymutations} ($M_{\tt := Y}$ and $M_{\tt +=Y}$), we can define the commutativity and subsumption relations by simple structural tests on the mutations being related: $\oraclecomm(M_{\tt := Y}, M_{\tt := Y})$, $\oraclecomm(M_{\tt +=Y}, M_{\tt +=Y'})$, and $\oraclesub(M, M_{\tt := Y})$ are the only relations that hold.  The identity mutation for this language is $M_{\tt +=0}$.
\end{example}

For more complex classes of mutations, this definition can be too strong.  Consequently, for the remainder of the paper, we will limit ourselves to {\em weak mutation languages}, where the relations $\oraclesub, \oraclecomm$ are conservative approximations.  If the relation holds then the corresponding property is guaranteed to hold, but not visa versa.  

Finally, we will define two notions of closure for a mutation language: First, a mutation language $\mutlang$ is closed over composition if the composition of two mutations $M, M' \in \mutlang$ is also in $\mutlang$.
$$\forall M, M' \in \mutlang : \exists M'' \equiv (M \circ M') \in \mutlang$$

Second, a mutation language $\mutlang$ is closed over binary operation $\theta : \tau \times \tau \mapsto \tau$ if there exists a mutation in $\mutlang$ that computes the result of applying $\theta$ to the output of mutations $M, M' \in \mutlang$.
$$\forall M, M' \in \mutlang : \exists M'' \in \mutlang : M''(x) \equiv (M(x)\theta M'(x))$$

\begin{example}
Our toy mutation language from Example \ref{ex:toylanguage} can be shown to be closed over composition, addition, and subtraction, but not multiplication.  The details of this proof are left to the reader.
\end{example}

We will use the two binary operations $\mutcmp$ and $\mutmerge_\theta$ to denote the result of combining two mutations by composition or by binary operation $\theta$ (respectively), for any mutation language closed over composition or $\theta$ (respectively).  Note that the existence of either function provably demonstrates the corresponding type of closure.


\subsection{Mutation Logs}
\label{sec:semantics-low:rworacle}

We now turn to our primary subject: logs.  Our goal in this section is to develop formalisms, first for the logs themselves, and second for reasoning about how the logs can be transformed, or rewritten, while preserving certain critical properties.

A {\em log} is a sequence of updates to an application's state, expressed as a numbered sequence of mutations: $M_1,\ldots, M_n$.

A log defines a corresponding sequence of application states: $v_0,\ldots,v_n$.  We obtain state $v_i$ by starting with a default state $v_0$, and applying mutations $M_1,\ldots,M_i$ in order.  In other words, for a mutation language closed under composition, $v_i$ is the result of composing the first $x$ mutations in the log.
$$v_i = (M_1\circ\ldots\circ M_i)(v_0)$$
We refer to the subscript of a state or mutaiton as its {\em timestamp} (i.e., $v_i$ and $M_i$ have timestamp $x$).  We define the {\em current state} of a log of size $n$ to be the state $v_n$.  The current state can be {\em recovered} from any intermediate state $v_i$ by applying the composition of all mutations after $x$.
$$v_n = (M_{x+1}\circ\ldots\circ M_n)(v_i)$$
Recovery is central to the design of Laasie.  A client can recover from a transient disconnection by replaying only those mutations that occurred while the client was disconnected, rather than forcing it to reload the full application state from scratch.

\subsubsection{Log Rewrites}

A log rewrite $\mathcal R$ is defined generally as an operation that transforms one sequence of mutations $M_1, \ldots, M_n$ into a new sequence $M_1', \ldots, M_{n'}'$.  

Because of our interest in recovery, we are interested in preserving a correspondence between timestamps in the pre- and post- rewritten states $v_i$ and $v_i'$ (respectively).  Consequently, we will assume that each pre-rewrite state corresponds to the post-rewrite state with the same timestamp.  Note that this limits us to size-preserving rewrites.  As we will soon see, this can be done without loss of generality.

We specifically consider two classes of size-preserving log rewrites: {\em delete} and {\em compose}.

\tinysection{Delete}
We can effect a size-preserving deletion rewrite by replacing the deleted mutation with the no-op identity operation ({\bf id}).  The rewrite $\mathcal R_{\tt del}(x)$, which deletes mutation $M_x$ is defined as
$$M_i' = \left\{\begin{array}{lcl} M_i & \ldots & i \neq x \\ \id & \ldots & i = 
x \end{array}\right.$$

\tinysection{Compose}
For a mutation language closed over composition, we can merge two mutations into the log into a single log entry.  The log size is preserved by inserting an {\bf id} mutation.  For reasons that will soon become clear, the composite mutation replaces the mutation with the higher timestamp, and the inserted {\bf id} replaces the mutation with the lower timestamp.  The rewrite $\mathcal R_{\tt cmp}(x, y)$, which merges mutations $M_x$ and $M_y$ is defined as
$$M_i' = \left\{\begin{array}{lcl} M_i & \ldots & i \not \in \{x,y\} \\ \id & \ldots & i = x \\ M_{x} \circ M_{y} & \ldots & i = y \end{array}\right.$$

\subsubsection{Rewrite Properties}

Now that we have defined log rewrites, we begin to consider what constitutes a legitimate log rewrite.  We define three correctness properties for log rewrites: {\em tail-correctness}, {\em recoverability}, and {\em $\vec t$-recoverability}.  We will also show how to use the subsumption and commutativity relations of a mutation language $\oraclesub$, $\oraclecomm$ to determine when these properties are guaranteed to be satisfied, independent of data, for a delete, compose, or commute rewrite.

\tinysection{Tail-Correctness} We start with the simplest of the log-rewrite properties.

\begin{definition}
A log rewrite is tail-correct if the current state $v_n$ of the log is identical to the current state $v_n'$ of the rewritten log.  That is:
$$(M_1 \circ \ldots \circ M_n)(v_0) = (M_1' \circ \ldots \circ M_n')(v_0)$$ 
\end{definition}

\begin{lemma}
\label{lem:correctness-del}
The rewrite $\mathcal R_{\tt del}(x)$ is tail-correct if $M_x$ is subsumed by the aggregate composition of all mutations following it: $\oraclesub(M_x, (M_{x+1} \circ M_{x+2} \circ \ldots \circ M_{n}))$.
\end{lemma}
\proof{The identity operation has no effect on the state, and can be inserted anywhere.  By subsumption, we have that 
$$M_{x} \circ \ldots \circ M_{n} \equiv M_{x+1} \circ \ldots \circ M_{n}$$
Thus, $v_n = v_n'$ \punto}

\begin{lemma}
\label{lem:correctness-com}
The rewrite $\mathcal R_{\tt cmp}(x,y)$ is tail-correct for any mutation language closed over composition if $M_x$ commutes with the aggregate composition of all mutations between it and $M_y$: $\oraclecomm(M_x, (M_{x+1}\circ\ldots\circ M_{y-1}))$
\end{lemma}
\proof{As before, identity has no effect on the state.  If $x = y-1$, then the merged mutations is equivalent to the separate mutations by Proposition \ref{compisassoc}.  Otherwise, by commutativity, we have that 
$$M_{x} \circ \ldots \circ M_{y-1} \equiv M_{x+1} \circ \ldots \circ M_{y-1} \circ M_{x}$$
Once $M_x$ and $M_y$ are adjacent, they can be merged just as before. \punto}

\begin{example}
Consider our toy mutation language from Example \ref{ex:toylanguage}.  From the subsumption relation $\oraclesub$, we can infer that it is tail-correct to delete any mutation preceding a replace mutation ($M_{\tt := Y}$).  

From the commutativity relation $\oraclecomm$, we can infer that it is tail-correct to merge any two mutations in an unbroken sequence of increment mutations ($M_{\tt += Y}$), or to merge a replace mutation with its immediate successor.  
\end{example}

\tinysection{Recoverability} Although tail-correctness provides a useful baseline for further discussions of log rewrites, it only takes a single state: the current state into consideration.  As such, it fails to capture any of the benefits of having a log in the first place.  We now consider a property that is strictly stronger than tail-correctness, and which allows us to reason about the possibility of recovery from any intermediate state.  We start with a per-timestamp notion of recoverability

\begin{definition}
A log rewrite is recoverable from timestamp $i$ (or equivalently state $v_i$) if the final state $v_n$ of the original log can be obtained by applying the sequence of rewritten mutations following timestamp $i$ to the state $v_i$, taken from the original log.
$$(M_1 \circ \ldots \circ M_n)(v_0) = (M_{i+1}' \circ \ldots \circ M_n')(v_{i})$$
Or equivalently (because $v_i$ is defined by the original log)
$$(M_1 \circ \ldots \circ M_n)(v_0) = (M_1 \circ\ldots\circ M_i \circ M_{i+1}' \circ \ldots \circ M_n')(v_0)$$
\end{definition}

\begin{definition}
A log rewrite is recoverable if it is recoverable from all timestamps in the log ({\em i.e.}, $i \in [0,n])$)
\end{definition}

Note that tail-correctness is the special case of recoverability from timestamp $0$.

\begin{lemma}
\label{lem:recover-del}
If the log rewrite $\mathcal R_{\tt del}(x)$ is tail-correct, it is recoverable
\end{lemma}

\proof{Recoverability from any state $v_i$ s.t. $i < x$ is equivalent to tail-correctness, because these states are unaffected by the rewrite.  Recoverability when $i \geq x$ is guaranteed always: The state $v_i$ being recovered from is taken before the rewrite, and mutations $M_{x+1}',\ldots,M_{n}'$ are identical to their pre-rewrite counterparts. \punto}

\medskip
This proof shows a tight coupling between correctness and recoverability, and illustrates an intriguing log partitioning.  If a rewrite only modifies mutations that fall within a fixed range, recoverability ``errors'' can only occur at states that fall within that same range.

\begin{proposition}
\label{correctintorecoverable}
Let $\mathcal R$ be a tail-correct log rewrite, which only alters log entries at timestamps in the range $[x,y]$.  Mutations outside of this range are unaffected by $\mathcal R$.  

$\mathcal R$ is recoverable iff it is recoverable from all states $v_i \in [x,y)$
\end{proposition}

\proof{The proof is identical to that of Lemma \ref{lem:recover-del}. \punto}

\begin{lemma}
\label{lem:recover-com}
The rewrite $\mathcal R_{\tt cmp}(x, y)$ is recoverable if it is correct, and if $M_x$ is idempotent: $\oraclesub(M_x,M_x)$
\end{lemma}

\proof{From the commutativity property required to show correctness, we have that $M_{x}\circ\ldots\circ M_{y-1} \equiv M_{x+1}\circ\ldots\circ M_{y-1}\circ M_{x}$.  For all $i \geq x$, state $v_i = (M_{x}\circ\ldots\circ M_{i})(v_{x-1})$.  Thus, $(M_{i+1}'\circ M_{y}')(v_i) \equiv (M_x\circ\ldots\circ M_x \circ M_y)(v_x)$.  By commutativity, we can rewrite this expression as $M_{x+1}\ldots\circ M_x\circ M_x \circ M_y$.  By idempotence, this is equivalent to the original rewritten expression, and by Proposition \ref{correctintorecoverable} the proof devolves to that of correctness.}

\begin{example}
Returning to the toy mutation language from Example \ref{ex:toylanguage}, we see that although it is tail-correct to merge any two increment mutations, it is not recoverable.  

Consider the log $\left(M_{\tt := 1}, M_{\tt += 2}, M_{\tt += 3}\right)$.  After applying the rewrite $\mathcal R_{\tt cmp}(2,3)$, we get $\left(M_{\tt := 1}, {\bf id}, M_{\tt += 5}\right)$.  After the rewrite, it is no longer possible to recover from state $v_2$ ($=3$), as the mutation $M_{\tt += 2}$ would effectively be applied twice.
\end{example}

\tinysection{$\vec t$-recoverability}  The intent of recoverability is to protect disconnected clients from reaching an inconsistent state when log entries are replayed.  However, to guarantee full recoverability, we must discard many potentially useful log rewrites.  In a practical setting, a server will not need to guarantee recoverability for all timestamps.

\begin{definition}
Given a set of timestamps $\vec t$, a log rewrite is $\vec t$-recoverable if it is recoverable from every $t \in \vec t$.
\end{definition}

By tracking when clients disconnect (regardless of whether or not the disconnection is transient), the server can identify ranges of log entries over which non-recoverable log rewrite can still be performed.

\begin{theorem}
Let $\mathcal R$ be a a tail-correct, but non-recoverable log rewrite, Let $[x, y]$ be the minimal range of timestamps affected by $\mathcal R$.  $\mathcal R$ is $\vec t$-recoverable iff $\left(\vec t \cap [x, y)\right) = \emptyset$.  
\end{theorem}

\proof{Follows from Proposition \ref{correctintorecoverable}}

\newcommand{\lbar}{\bar\mutlang}
\newcommand{\collbar}{\coll\mutlang(\lbar)}
\newcommand{\idmut}{\mathcal{ID}}

\section{Reducing \barql to LBar}
\label{sec:reduction}

We now apply the principles of mutation languages to \barql by constructing a weak mutation language $\lbar$ (LBar) built around \barql.  Roughly speaking, this mutation language allows a single monolithic \barql query to be subdivided into a set of disjoint operations, each applied to a specific point in the path hierarchy.  This allows us to easily identify the {\em write dependencies} of a \barql query at their finest granularity.

We then transform each subdivided operation into a delta form, with a \barql query that computes a {\em delta value} and a merge operator, a binary function that defines how the delta value is to be merged with the prior state.  This {\em update operator} simplifies the task of determining commutativity and subsumption at a fine granularity.

We also identify the set of points in the path hierarchy that each query reads from.  This set of points forms the set of {\em read dependencies} of the query.  

Finally, we use the sets of write dependencies, read dependencies, and update operators to efficiently compute the commutativity and subsumption relations $\oraclecomm, \oraclesub$ for a \barql query. 

\subsection{LBar}
The typesystem of $\lbar$ is identical to that of \barql.  To recap: values can be of any primitive type, or a collection, which is a mapping from key names of abstract type $k$ to values.  Collections can be organized into a hierarchy.  We use $\phi$ to denote an ordered sequence of key names that defines a {\em path} through the collection herarchy.

Point mutations form the basis of $\lbar$, and express updates to individual paths in a \barql hierarchy.  A {\em point mutation} is a 3-tuple $\tuple{\phi, Q, (\theta\ |\ \emptyset)}$, where $\phi$ is the path being updated and $Q$ is a \barql expression that computes an {\em update delta} based on the prior state.  Every point mutation is annotated with either a binary operation $\theta$, or the {\em overwrite} annotation $\emptyset$.  The annotation indicates the {\em combinator} used to merge delta value with the original.  

We say that two point mutations are path-disjoint if neither point mutation's path is a prefix of the other's.  A {\em full mutation} in $\lbar$ is a set of pairwise path-disjoint point mutations, which it applies to the state in parallel; The prior state for all point mutations in the set is defined uniformly to be the prior state for the full mutation.  Thus, all point mutations are guaranteed to be isolated in the traditional database sense.

As a shorthand, we will use $\omega(\mut)$ to denote the write set of a full mutation $\mut$, the set of all paths of point mutations in the full mutations:

$$\omega(\mut) = \{\phi\ |\ \tuple{\phi, Q, \theta} \in \mut\}$$

We will also use the shorthand $\mut[\phi]$ to denote the point mutation applied to path $\phi$ for all $\phi \in \omega(\mut)$.

\subsection{Reduction Algorithm}
We now present an iterative process for transforming \barql expressions into $\lbar$ form.  This process begins by creating a full mutation consisting of a single point mutation $\{\tuple{[], Q, \emptyset}\}$.  

The algorithm repeatedly selects an arbitrary point mutation in the set and tries (1) to subdivide point-mutations in this set into finer-grained mutations, and (2) to replace overwrite annotations by extracting binary operations from the point-mutation's query.  This process proceeds up to a fixed point.

\tinysection{Operator-Extraction}
In their simplest incarnations, both transformations are applied to point queries of the same general form:
$$\tuple{\phi, (\id.\phi\ \theta\ Q'), \emptyset}$$
For a $\theta$ that is commutative and associative, any query with a $\id.\phi$ term can be commuted to the front.  

In this expression, $Q'$ effectively expresses the delta of the point update, while $\theta$ combines it with the original value $\id.\phi$.  Consequently, $\theta$ becomes the new combinator, and $Q'$ becomes the new update delta.

\tinysection{Key-Extraction}
The merge operator ($\merge$) is associative (but not commutative).  As with binary operators on primitive type, we can compute an update delta of expressions that derive from $\id$.  We start by identifying the {\em change set} of the original query.  We start from a point update of the form:
$$\tuple{\phi, Q, \emptyset}$$
If a query $Q$ returns a value of collection type, its change set $\delta(Q)$ is computed as follows:

\begin{itemize}
\item $\delta(\emptyset) = \emptyset$
\item $\delta(\{k := Q'\} = \{k\}$
\item $\delta(\id.\phi) = \{*\}$
\item $\delta(\id.\phi') = $ This point mutation can not be subdivided.
\item $\delta(\map{\ldots}{Q'}) = \delta(Q')$
\item $\delta(\filter{\ldots}{Q'}) = \delta(Q')$
\item $\delta(Q' \merge Q'') = \delta(Q') \cup \delta(Q'')$
\item $\delta(\ifq{\ldots}{Q'}{Q''}) = \delta(Q') \cup \delta(Q'')$
\item $\delta(Q' \circ Q'') = \delta{Q''[\id / Q']}$
\end{itemize}
The key $*$ is a special key that refers to all keys in the input query input.  This special key is treated as a distinct key in the changeset computation.  If it is in the changest for a delta query ($* \in \delta(Q)$), the point mutation modifies the original value (instead of overwriting it), and can be subdivided further as follows.

We begin by generating a delta computation $\Delta_k(Q)$ for each subkey $k$ in the changeset.  This includes a delta computation for the special key $*$, which will be applied to all keys in the input that are not explicitly present in the changeset.

\begin{itemize}
\item $\Delta_k(\emptyset) = \Null$
\item $\Delta_k(\{k := Q'\}) = Q'$
\item $\Delta_k(\{k' := Q'\}) = \Null$
\item $\Delta_k(\id.\phi) = \id.\phi.k$
\item $\Delta_k(\map{Q''}{Q'}) = \Delta_k(Q') \circ Q''$
\item $\Delta_k(\filter{Q''}{Q'}) = $\\
$\ifq{\Delta_k(Q') \circ Q''}{\Delta_k(Q')}{\Null}$
\item $\Delta_k(Q' \merge Q'') = $\\
$\ifq{\Delta_k(Q'') \neq \Null}{\Delta_k(Q'')}{\Delta_k(Q')}$
\item $\Delta_k(\ifq{\Delta_k(Q)}{Q'}{Q''}) = $\\
$\ifq{\Delta_k(Q)}{\Delta_k(Q')}{\Delta_k(Q'')})$
\item $\Delta_k(Q' \circ Q'') = \Delta_k{Q''[\id / Q']}$
\end{itemize}

The resulting expression can be simplified by partial evaluation.  In many cases, it will be possible to eliminate operations over $\Null$ values.  The result is a set of point mutations, one for each key $k$ in the changeset, including the special key $*$.  Once again, $*$ applies to all children at $\phi$ except those explicitly defined (by being present in the changeset).  The resulting set of point mutations is thus defined as
$$\{\tuple{\phi.k, \Delta_k(Q), \emptyset}\ |\ k \in \delta(Q) \wedge (\Delta_k(Q) \neq \id.\phi.k)\}$$
Note that we explicitly exclude the identity mutation, as this is effectively a no-op.  

\subsection{Read Dependencies}

We compute the read dependencies of a \barql query by first defining a read-normal form for \barql.  We call a query of the form $\id.k_1.k_2.(\ldots).k_n$ a {\em point read} at path $\phi = k_1.k_2.(\ldots).k_n$.  A query is in read-normal form if the subscript operator appears only in point reads, or is applied to the special key $tmp$, defined below.  As we now show, any valid query can be transformed into read-normal form:

\begin{itemize}
\item $(Q \merge Q').k \mapsto \ifq{Q'.k \neq \Null}{Q'.k}{Q.k}$
\item $(\map{Q'}{Q}).k \mapsto Q.k \circ Q'$
\item $(\agg{\merge}{Q}).k \mapsto $\\
$(\agg{\merge}{\map{\{tmp := \id.k\}}{Q}}).tmp$
\item $(\filter{Q'}{Q}).k \mapsto \ifq{Q.k \circ Q'}{Q.k}{\Null}$
\item $(\ifq{Q}{Q'}{Q''}).k \mapsto \ifq{Q}{Q'.k}{Q''.k}$
\item $(Q \circ Q').k \mapsto Q \circ (Q'.k)$
\item $\emptyset.k \mapsto \Null$
\end{itemize}

Given a query $Q$ in read-normal form, we can compute the readset of the query $\rho(Q)$ as follows:

\begin{itemize}
\item $\rho(\id.\phi) = \{\phi\}$
\item $\rho(Q \merge Q') = \rho(Q) \cup \rho(Q')$
\item $\rho(\map{Q'}{Q}) = \rho(Q)$\footnote{This is a conservative approximation.}
\item $\rho(Q {\bf op}_{\theta} Q') = \rho(Q) \cup \rho(Q')$
\item $\rho(\agg{\theta | \merge}{Q}) = \rho(Q)$
\item $\rho(\filter{Q'}{Q}) = \rho(Q)$
\item $\rho(\ifq{Q}{Q'}{Q''}) = \rho(Q) \cup \rho(Q') \cup \rho(Q'')$
\item $\rho(Q \circ Q') = \rho{Q'[\id / Q]}$
\item $\rho(c | \Null | \emptyset) = \emptyset$
\end{itemize}

\subsection{Subsumption and Commutativity}
We are now ready to complete the definition of the mutation language 4-tuple for $\lbar$ by defining a conservative approximation of the subsumption and commutativity relations.  

\tinysection{Subsumption}
A path $\phi$ is subsumed by a full mutation $\mut$ if it or one of its ancestors is {\em overwritten} by $\mut$, and neither $\phi$, nor any of its ancestors or descendents appear in the read set of $\mut$.  Abusing syntax, we write this as:
\begin{multline*}
\oraclesub(\phi, \mut) \equiv \\
(\exists Q, \phi' \in \omega(\mut) : (\phi' \sqsubseteq \phi) \wedge (\mut[\phi'] = \tuple{\phi', Q, \emptyset}))\\
\wedge (\not \exists \phi' \in \rho(\mut) : (\phi' \sqsubseteq \phi)\vee(\phi \sqsubseteq \phi'))
\end{multline*}

Here, $\sqsubseteq$ denotes the ancestor of relation.

A mutation $\mut$ is subsumed by $\mut'$ if all paths in the write set of $\mut$ are subsumed by $\mut'$:
$$\oraclesub(\mut, \mut') \equiv \forall \phi \in \omega(\mut') : \oraclesub(\phi, \mut')$$

\tinysection{Commutativity}
Two point mutations applied to the same path $\phi$, $\tuple{\phi, Q, \theta}$ and $\tuple{\phi, Q', \theta'}$ commute iff $\theta$ commutes with $\theta'$.  Two point mutations applied to different paths, $\tuple{\phi, Q, \theta}$ and $\tuple{\phi', Q', \theta'}$ commute iff each of the following conditions holds: (1) $\phi$ is neither an ancestor, nor descendant of $\phi'$, (2) $\phi$ is neither an ancestor, nor descendant of a path in the read set $\rho(Q')$, and (3) $\phi'$ is neither an ancestor, nor descendant of a path in the read set $\rho(Q)$.

Two full mutations commute, if all pairs of point mutations commute.  Again, abusing syntax:
$$\oraclecomm(\mut, \mut') \equiv \forall m \in \mut, m' \in \mut' : \oraclecomm(m, m')$$


\section{Related Work}
\label{sec:related}
There has been much work focused on the formalization of query languages and database models~\cite{Abiteboul:1995:PLM:615232.615234,Abiteboul:1987:IFS:32204.32205,Majkic:2010:KCD:1868384.1868389}.
Much of this work is based on monad algebra, Lawvere theories, and universal algebra~\cite{Jaeschke:1982:RAN:588111.588133,Balan:2010:COA:1841982.1842045,Adamek:2011:CAC:2040096.2040103,Hyland:2007:CTU:1230168.1230590}.
Manes {\em et al.}~\cite{Manes:1998:ICC:967301.967303} showed how
to implement collection classes using monads.
Cluet~\cite{Cluet:1990:RAB:96448.96489} is an
algebra based query language for an object-oriented database system.
Our work is based on the same fundamental theories. In the following we compare our work to previous results.

\tinysection{Languages for Transforming Hierarchical Data}
There has been considerable work~\cite{buneman1995principles,abiteboul1997lorel,Abiteboul:1995:PLM:615232.615234,abiteboul1986non} on the transformation of hierarchical data.  Two approaches have become dominant in this area: Nested Relational Calculus~\cite{roth1988extended} and the Monad Algebra~\cite{lellahi1997calculus}.  Our own approach is closely based on the latter, adapted for use with labeled sets, and with the intentional exclusion of the superlinear time complexity pairwith operator (or equivalently, the cartesian cross-product).

\tinysection{Semistructured Data}
Also closely related is work on managing semistructured data~\cite{buneman1997semistructured}.  The vast majority of recent efforts in this area have been on querying and transforming XML data.  One formalization by Koch~\cite{Koch:2006:CNX:1189769.1189771} is also closely based on Monad Algebra.  Work by Cheney follows a similar vein, in particular (F)LUX~\cite{cheney2007lux,cheney2008flux}, a functional language for XML updates.  In \cite{benedikt2009semantics}, Benedikt and Cheney present a formalism for synthesizing the output schema of XML transformations, similar to our notion of the compositional compatibility of mutations.  More recently, there has also been interest in querying lighter-weight semistructured data representations like JSON\cite{beyer2011jaql,beyer2009jaql}.

\tinysection{Algebraic Properties of State Updates}
The distributed systems community has identified a number of algebraic properties of state mutations that are useful in distributed concurrency control.  Commutativity of updates has been explored extensively~\cite{Weihl:1988kj,Shapiro:2007wf}, but the typical assumption is that a domain-specific commutativity oracle is available, such as for edits to textual data~\cite{Shapiro:2007wf,Oster:2006ee}.  
Our notion of subsumption is quite similar to the Badrinath and Ramamritham~\cite{Badrinath:1990dp}'s recoverability property.  Unlike subsumption, this property is defined in terms of observable side-effects rather than state, but is otherwise identical.  Like prior work on commutativity, they assume that a domain-specific oracle has been provided.  
Several efforts have been made to understand domain-specific reconciliation strategies.  Feldman {\em et al.}'s Operational Transforms~\cite{Feldman:wl} are analogous to our our mutation languages, but assume that domain-specific operations analogous to our merge operation are available.  
Perhaps the closest effort to our own has been Preguica {\em et al.}'s IceCube~\cite{Shapiro:2003uw}, and Edwards {\em et al.}'s Bayou~\cite{Edwards:1997wl}, each of which exploit a range of specific algebraic properties of updates to distributed state.  However, both systems must be explicitly adapted to specific application domains by the construction of domain-specific property oracles, or by mapping the application's behavior down to a trivial update language.
To the best of our knowledge, none of these areas have been explored in the context of a non-trivial state update language.

\tinysection{Update Sequencing}
The use of distributed logs and publish/subscribe to apply a canonical order to updates has also been explored extensively by the distributed systems and database communities.  Ellis {\em et al.} noted the relevance of sequencing to distributed concurrency control~\cite{Ellis:1989ju}.  Eugster {\em et al.} identified the usefulness of sequencing updates to distributed collection types~\cite{Eugster:2000tf}.  Domain specific applications of similar ideas can be found in work by Ostrowski and Birman~\cite{Ostrowski:2010bh}, Weatherspoon {\em et al.}~\cite{Weatherspoon:2007er}, 
and others.

\tinysection{Intent-Based Updates}
The use of intent-based ({\em i.e.}, operational) updates appears frequently in database literature, especially in the context of distributed databases, where it is used to reduce communication overhead.  Two concrete examples are Ceri and Widom's Starburst~\cite{Ceri:1992wo}, and Chang {\em et al.}'s BigTable~\cite{chang2008bigtable}.





\bibliographystyle{plain}
\small
\bibliography{barql}

\end{document}
